\documentstyle[preprint,aps]{revtex}  
\begin{document}

\draft

\title{A simple solution for the flavor question}

\author{Felice Pisano}

\address{Instituto de F\'{\i}sica Te\'orica, 
         Universidade Estadual Paulista\\ 
         Rua Pamplona 145 - 01405-000 - S\~ao Paulo, S.P. \\
          Brazil}


\maketitle

\begin{abstract}
We consider a simple way for solving the flavor question 
by embedding the three-familiy Standard Model in a semisimple gauge 
group extending minimally the weak isospin factor. Quantum chiral anomalies   
between families of fermions 
cancel with a matching of the number of families and the number of color   
degrees of freedom. Our demonstration shows how the theory    
leads to determination of families structure when the Standard Model is the 
input at low energies. The new physics is limited to start below a 
few TeVs within the reach of the next generation colliders. 
\end{abstract}

In the Standard Model~\cite{um} the fundamental fermions come in 
families. In writing down the theory one may start by first 
introducing just one family, then one may repeat the same procedure   
by introducing copies of the first family. Why do quarks and leptons 
come in repetitive structures (families)? How many families are 
there? How to understand the inter-relation between families? 
These are the central issues of the weak interaction physics known 
as the flavor question or the family problem. Nowhere in physics  
this question is replied~\cite{glashow}. 
One of the most important experimental results in the   
past few years has been the determination of the number of these  
families within the framework of the Standard Model. In the 
minimal electroweak model the number of families is given by the  
number of the neutrino species which are all massless, by 
definition. The number of families is then computed from the 
invisible width of the $Z^0$, 
\begin{equation}
\Gamma_{\rm inv}\equiv 
\Gamma_{Z^0} - (\Gamma_h + \sum_l \Gamma_l)   
\label{eq1}
\end{equation} 
where $\Gamma_{Z^0}$ denotes the total width, the subscript $h$ refers 
to hadrons and $\Gamma_l$, $l = \{e,\,\mu,\,\tau\}$, is the width of 
the $Z^0$ decay into an $l\bar l$ pair. If $\Gamma_\nu$ is the  
theoretical width for just one massless neutrino, the number 
of families is  
\begin{equation} 
N_{\rm fam} = N_\nu = \frac{\Gamma_{\rm inv}}{\Gamma_\nu}
\label{eqdois}
\end{equation}
and recent results give a value very close to three~\cite{dois}  
$$
N_{\rm fam} = 2.99\pm 0.03
$$
but we don't understand why the number of standard families is three. 
The answer to the flavor question may require a radical change in our  
approaches. It could be that the underlying objects are strings and 
all the low energy phenomena will be determined by physics at the 
Planck scale. Grand Unified Theories (GUT) have had a major impact on 
both Cosmology and Astrophysics; for Cosmology they led to the  
inflationary scenario, while for Astrophysics supernova neutrinos  
were first observed in proton-decay detectors. It remains for GUTs  
to have impact directly on particle physics itself~\cite{doisA}. GUTs     
cannot explain the presence of fermion families. On the other side,   
supersymmetry for the time being is an answer in search of question 
to be replied.    
It doesn't explain the existence of any known particle or symmetry.  
Some traditional approaches to the problem such as GUTs, monopoles   
and higher dimensions introduce quite speculative pieces of new 
physics at high and experimentally inaccessible energies. Some years  
ago there were hopes that the number of families could be computed 
from first principles such as geometry of compactified manifolds but  
these hopes did not materialize. The Standard Model works so well, 
that there is, at present, no experimental evidence for new physics   
beyond the Standard Model. Of course, this does not mean that there is   
no new physics. 
\par
We wish to suggest here that some very fundamental aspects of the 
Standard Model, in particular the flavor question, might be understood    
by embedding the three-family version in a Yang-Mills theory with the 
gauge semisimple group~\cite{tres,quatro} 
$$
G_{331}\equiv \mbox{SU(3)}_C\otimes\mbox{SU(3)}_L\otimes\mbox{U(1)}_N
$$
with a corresponding enlargement of the quark representations. In  
particular, the number of families will be related by anomaly 
cancellation to the number of quark colors. In the 
$$
G_{321}\equiv \mbox{SU(3)}_C\otimes\mbox{SU(2)}_L\otimes\mbox{U(1)}_Y
$$
low-energy limit all three families appear similar and cancel anomalies   
separately. The $G_{331}$ model is a dilepton gauge theory which is 
chiral and has nontrivial anomaly cancellation. This novel method of 
anomaly cancellation requires that at least one family transforms 
differently from the others, thus breaking generation universality.   
Unlike the $G_{321}$ Standard Model, where anomalies cancel family by 
family, anomalies in the $G_{331}$ model only cancel when all three 
families are taken together. With this meaning we present here the 
simplest solution for the flavor question just enlarging the SU(2)$_L$  
weak isospin group to SU(3)$_L$. This does not expalin why 
$N_{\rm fam} > 1$ for the number of families but is sufficiently 
impressive to suggest that $N_{\rm fam} = 3$ may be explicable by 
anomaly cancellation in the simplest gauge extension of the Standard 
Model with a very particular representation content. The electroweak 
gauge group extension from SU(2) to SU(3) will add five gauge bosons. 
The adjoint gauge octet of SU(3) breaks into 
${\bf 8} = {\bf 3} + ({\bf 2} + {\bf 2}) + {\bf 1}$ under SU(2).  
The ${\bf 1}$ is a $Z^\prime$ and the two doublets are readily 
identifiable from the leptonic triplet or antitriplet  
$(\nu_l,\,l^-,\,l^+)$ as dilepton gauge bosons $(U^{--},V^-)$ and  
$(U^{++},V^+)$. Such dileptons appeared first in stable-proton 
GUT~\cite{cinco} but there the fermions were non-chiral and one 
needed to invoke mirror fermions; this is precisely what is avoided 
in the $G_{331}$  model. Contrary to the GUT case, there is 
no ``grand desert'' if $G_{331}$ models are realized in nature and 
new physics could arise at not too high energies, say in the TeV 
range~\cite{seis}. 
\par
We start with the way the electric charge operator ${\cal Q}$ is  
embedded in the neutral generators of the SU(3)$_L$ group. The  
fermion contents depend on the electric charge operator 
\begin{equation} 
{\cal Q} = \frac{1}{2}(\lambda^L_3 + \xi \lambda^L_8) + N
\label{uno}
\end{equation}
where $\lambda^L_{3,8}$ are the neutral generators of SU(3)$_L$, 
$\xi$ is the embedding parameter and $N$ is the U(1)$_N$ charge  
proportional to the unit matrix. The SU(3)$_L$ generators are  
normalized as Tr$(\lambda^L_a\lambda^L_b) = 2\delta_{ab}$;  
$a,\,b = 1,2,...,8$. In the $G_{331}$ models with lepton charges 
$0,\pm 1$ there is always a set of families transforming as 
$({\bf 1},{\bf 3}, 0)$ under the gauge group. In these families 
there is charge quantization in the sense of GUTs; the electric 
charge operator is a linear combination of the simple group 
generators. 
\par
In the $\xi = -\sqrt 3$ model~\cite{tres} three families of 
leptons belong to representation 
\begin{equation}
\psi_{lL}\equiv 
\left ( 
\begin{array}{c}
\nu_l \\ 
l \\
l^c 
\end{array}
\right )_L 
\sim ({\bf 1},{\bf 3},N_{\psi_{lL}} = 0); \quad 
l = {e,\,\mu,\,\tau }
\label{eq2}
\end{equation}
where $l^c = C \bar l^T$ and $C$ being the charge conjugation matrix. 
The right-handed neutrinos may be included in the theory if  
desired~\cite{seteA}.  
A result of this embedding is that there are no new leptons in the 
$G_{331}$ model. While all three lepton families are treated 
identically, anomaly cancellation requires that one of the quark 
families transforms differently from the other two. In particular,  
canceling the pure SU(3)$_L$ anomaly requires that there are the same 
number of triplets and antitriplets. Taking into account the three 
quark color degrees of freedom we must introduce the multiplets of 
chiral quarks 
\begin{equation}
Q_{1L} \equiv 
\left (
\begin{array}{c}
u \\
d \\
J
\end{array}
\right )_L
\sim ({\bf 3},{\bf 3},N_{Q_{1L}}); 
\quad
Q_{2,3L}\equiv 
\left (
\begin{array}{c}
j_1,\,j_2 \\
c,\, t \\
s,\, b
\end{array}
\right )_L 
\sim ({\bf 3},{\bf 3^*},N_{Q_{2,3L}})
\label{eq3}
\end{equation}
with the respective right-handed fields in SU(3)$_L$ singlets, 
\begin{eqnarray}
u_R & \sim & ({\bf 3},{\bf 1},N_{u_R}), 
\quad
c_R \sim ({\bf 3},{\bf 1},N_{c_R}), 
\quad
t_R \sim ({\bf 3},{\bf 1}, N_{t_R}); \nonumber \\
d_R & \sim & ({\bf 3},{\bf 1},N_{d_R}), 
\quad
s_R \sim ({\bf 3},{\bf 1},N_{s_R}), 
\quad 
b_R \sim ({\bf 3},{\bf 1},N_{b_R}),  
\label{eqquatro}
\end{eqnarray}
and the exotic quarks 
\begin{equation}
J_R \sim ({\bf 3},{\bf 1},N_{J_R}), 
\quad
j_{1R} \sim ({\bf 3},{\bf 1},N_{j_{1R}}), 
\quad 
j_{2R} \sim ({\bf 3},{\bf 1},N_{j_{2R}}) 
\label{eqcinco}
\end{equation}
where we have suppressed the color index. We are dealing with a gauge 
theory of chiral fermions. There are two quite distinct ways in which  
the $G_{331}$ model establish the inter-relation between fermion 
families. Firstly, there are a set of constraints wich follow from 
the consistency of the theory at the classical level, such as the 
requirement that the Lagrangian be gauge invariant, while there are 
other constraints which follow from the consistency of the theory 
at the quantum level which are the anomaly cancellation conditions. 
Anomalies imply the loss of a classical symmetry in the quantum 
theory~\cite{sete}. For chiral gauge theories in four dimensions 
our basic tool will be freedom from the triangle perturbative 
chiral gauge anomaly which must be canceled to avoid the breakdown 
of gauge invariance and the renormalizability of the theory. Of 
course, it is clear that anomalies alone cannot lead to a definite 
theory without some way to specify the underlying chiral fermions 
and some knowledge of the gauge symmetry that is responsible for 
the dynamics. 
\par 
Let us first obtain the classical constraints. In 
order to generate Yukawa couplings we introduce the minimal set of 
scalar fields SU(3)$_L$ triplets 
$\eta \sim ({\bf 1},{\bf 3},N_\eta)$, 
$\rho \sim ({\bf 1},{\bf 3},N_\rho)$, 
and $\chi \sim ({\bf 1},{\bf 3},N_\chi)$. The Yukawa Lagrangian, 
without considering the mixed terms between quarks is 
\begin{eqnarray}
-{\cal L}^Y_Q & = & \bar{Q}_{1L}(G_u u_R \eta + G_d d_R \rho + 
                     G_J J_R \chi) 
               +  (G_c \bar{Q}_{2L} c_R + G_t \bar{Q}_{3L} t_R)\rho^* 
              \nonumber \\
              & + & (G_s \bar{Q}_{2L} s_R + G_b \bar{Q}_{3L} b_R)\eta^* 
                +  (G_{j_1} \bar{Q}_{2L} j_{1R} + 
                     G_{j_2} \bar{Q}_{3L} j_{2R})\chi^* + \mbox{H.c.}
\label{eqseis}
\end{eqnarray}
where all fields are weak eigenstates and $\eta^*$, $\rho^*$, $\chi^*$  
denote the respective antitriplets~\cite{oito}. The requirement of 
gauge invariance leads to the classical constraints   
\begin{eqnarray}\label{789}
N_{Q_{1L}} - N_{u_R} & = & N_\eta \nonumber \\
N_{Q_{1L}} - N_{d_R} & = & N_\rho \\  
N_{Q_{1L}} - N_{J_R} & = & N_\chi \nonumber
\end{eqnarray}
for the first family and 
\begin{eqnarray}\label{101112}
N_{Q_{2L}} - N_{j_{1R}} & = & N_{\chi^*} \nonumber \\
N_{Q_{2L}} - N_{c_R} & = & N_{\rho^*} \\
N_{Q_{2L}} - N_{s_R} & = & N_{\eta^*} \nonumber 
\end{eqnarray}
for the second family. The constraints for the third family are 
obtained from those of the second family making the replacements  
$Q_{2L}\rightarrow Q_{3L}$, $j_{1R}\rightarrow j_{2R}$, 
$c_R\rightarrow t_R$, and $s_R\rightarrow b_R$. The above equations 
with $N_{\eta^*} = - N_\eta$, $N_{\rho^*} = - N_\rho$, and 
$N_{\chi^*} = - N_\chi$ imply 
\begin{eqnarray}\label{felice}
N_{Q_{1L}} + N_{Q_{2L}} & = & N_{u_R} + N_{s_R} \nonumber \\
N_{Q_{1L}} + N_{Q_{2L}} & = & N_{d_R} + N_{c_R} \\ 
N_{Q_{1L}} + N_{Q_{2L}} & = & N_{j_{1R}} + N_{J_R} \nonumber
\end{eqnarray}
constraining the first and second families and 
\begin{eqnarray}\label{pisano}
N_{Q_{2L}} - N_{Q_{3L}} & = & N_{j_{1R}} - N_{j_{2R}} \nonumber \\
N_{Q_{2L}} - N_{Q_{3L}} & = & N_{c_R} - N_{t_R} \\
N_{Q_{2L}} - N_{Q_{3L}} & = & N_{s_R} - N_{b_R} \nonumber 
\end{eqnarray}
which relates the second and third families. This step illustrates   
how the Lagrangian is used as the primary source of constraints. 
\par
Let us now consider the quantum constraints. It will be sufficient 
to consider only anomalies which contain U(1)$_N$ factors 
\newpage
$$
\mbox{Tr}[\mbox{SU(3)}_C]^2[\mbox{U(1)}_N] = 0:
$$
\begin{eqnarray}
& 3 & (N_{Q_{1L}} + N_{Q_{2L}} + N_{Q_{3L}}) - N_{u_R} - N_{c_R} - N_{t_R} 
\nonumber \\
& - & N_{d_R} - N_{s_R} - N_{b_R} - N_{J_R} - N_{j_{1R}} - N_{j_{2R}} = 0
\label{eq19}
\end{eqnarray}
$$
\mbox{Tr}[\mbox{SU(3)}_L]^2[\mbox{U(1)}_N] = 0:
$$
\begin{equation}
3\,\,(N_{Q_{1L}} + N_{Q_{2L}} + N_{Q_{3L}}) + 
N_{\psi_{eL}} + N_{\psi_{\mu L}} + N_{\psi_{\tau L}} = 0
\label{eq20}
\end{equation}
$$
\mbox{Tr}[\mbox{U(1)}_N]^3 = 0:
$$
\begin{eqnarray}
& 3 & (N^3_{Q_{1L}} + N^3_{Q_{2L}} + N^3_{Q_{3L}}) - N^3_{u_R} - N^3_{c_R} 
- N^3_{t_R} \nonumber \\ 
& - & N^3_{d_R} - N^3_{s_R} - N^3_{b_R} - N^3_{J_R} - N^3_{j_{1R}} 
- N^3_{j_{2R}} \nonumber \\
& + & N^3_{\psi_{e L}} + N^3_{\psi_{\mu L}} + N^3_{\psi_{\tau L}} = 0   
\label{eq21}
\end{eqnarray}
$$
\mbox{Tr}[\mbox{graviton}]^2[\mbox{U(1)}_N] = 0:
$$
\begin{eqnarray}
& 3 & (N_{Q_{1L}} + N_{Q_{2L}} + N_{Q_{3L}}) - N_{u_R} - N_{c_R} - N_{t_R} 
\nonumber \\
& - & N_{d_R} - N_{s_R} - N_{b_R} - N_{J_R} - N_{j_{1R}} - N_{j_{2R}} 
\nonumber \\
& + & N_{\psi_{e L}} + N_{\psi_{\mu L}} + N_{\psi_{\tau L}} = 0 
\label{eq22}
\end{eqnarray}
where the first three anomalies are the familiar triangle gauge-anomalies 
and the last condition in Eq. (\ref{eq22}) is a little more speculative 
in that it arises from a triangle graph with two external gravitons   
and one $G_{331}$ gauge boson. Whatever the correct quantum gravity 
theory is, the ``mixed gauge-gravitational''~\cite{nove} anomaly 
must be cancelled for consistency. If one believes in quantum gravity,  
then one may also wish to impose the requirement that the mixed 
gauge-gravitational anomaly cancel. Notice that in 
contrast to the minimal    
Standard Model, the classical and the quantum constraints enclose  
all three families of fermions. As it was said before, the quark 
representations in Eqs. (\ref{eq3}) - (\ref{eqquatro}) are symmetry 
eigenstates; that is, they are related to the mass eigenstates 
by Cabibbo-like angles. As we have one triplet and two antitriplets,  
it should be expected that flavor-changing neutral currents exist.  
However when we determine the neutral currents explicitly we find 
that all of them, for the same charge sector, have equal factors 
and the Glashow-Iliopoulos-Maiani~\cite{dez} cancellation is 
automatic in neutral currents coupled to $Z^0$~\cite{tres,quatro}.  
Although each family is anomalous, this type of construction is 
only anomaly-free when the number of families is divisible by the 
number of colors. Thus three families are singled out as the 
simplest nontrivial anomaly-free $G_{331}$ model.
\par
The flavor question of the Standard Model might be understood by 
embedding the three family version in the $G_{331}$ group with a 
corresponding enlargement of the quark representations. In the 
$G_{331}$ low-energy limit all three families appear similarly and 
cancel anomalies separately. By matching the coupling constants 
at the $G_{331}$ symmetry breaking an upper limit on the   
symmetry-breaking scale of a few TeVs can be placed by the 
requirement that $\sin^2\theta_W < 1/4$, implyng that the physics 
associated with the $(U^{\pm\pm}, V^\pm)$ dilepton gauge bosons, 
the additional $Z^\prime$ neutral gauge boson, and the $J$, 
$j_{1,2}$ exotic quarks will be accessible to the next generation of 
colliders~\cite{seis,onze}. The Standard Model is the effective 
low energy theory of the $G_{331}$ model and it enjoys considerable  
support from experiment. As such we can take it to be a safe input  
to $G_{331}$. According to Eq. (\ref{eq2}) we have 
directly $N_{\psi_{lL}} = 0$ for   
any leptonic family $l = {e,\,\mu,\,\tau}$. Let us set the following    
notation 
\begin{eqnarray}
N_{u_R} & = & N_{c_R} = N_{t_R} \equiv N_{U_R}, \\ 
N_{d_R} & = & N_{s_R} = N_{b_R} \equiv N_{D_R}
\label{2324}
\end{eqnarray}
and from the constraints given in Eqs. (\ref{pisano})  we obtain 
the following two conditions 
\begin{equation}
N_{Q_{2L}} = N_{Q_{3L}} \equiv N_{Q_{\alpha L}}, \quad \alpha = 2,3;
\label{25}
\end{equation}
and
\begin{equation}
N_{j_{1R}} = N_{j_{2R}} \equiv N_{j_R}.
\label{26}
\end{equation}
Thus we write the quantum constraints of 
Eqs. (\ref{eq19}) - (\ref{eq22}) in the concise form 
\newpage  
$$
\mbox{Tr}[\mbox{SU(3)}_C]^2[\mbox{U(1)}_N] = 0:
$$  
\begin{equation}
3(N_{Q_{1L}} + 2 N_{Q_{\alpha L}}) - 3(N_{U_R} + N_{D_R}) - N_{J_R} 
- 2 N_{j_R} = 0
\label{27}
\end{equation}
$$
\mbox{Tr}[\mbox{SU(3)}_L]^2[\mbox{U(1)}_N] = 0:
$$
\begin{equation}
3 (N_{Q_{1L}} + 2 N_{Q_{\alpha L}}) = 0
\label{28}
\end{equation}
$$
\mbox{Tr}[\mbox{U(1)}_N]^3 = 0:
$$
\begin{equation}
3 (N^3_{Q_{1L}} + 2 N^3_{Q_{\alpha L}}) - 3 (N^3_{U_R} + N^3_{D_R}) 
- N^3_{J_R} - 2 N^3_{j_R} = 0
\label{29}
\end{equation}
and the mixed gravitational-gauge constraint coincides with the 
$[\mbox{SU(3)}_C]^2[\mbox{U(1)}_N]$ anomaly. In the new notation 
the classical constraints given in Eqs. (\ref{felice})             
becomes 
\begin{eqnarray}
N_{Q_{1L}} + N_{Q_{2L}} & = & N_{U_R} + N_{D_R}, \nonumber \\
N_{Q_{1L}} + N_{Q_{2L}} & = & N_{j_R} + N_{J_R}.
\label{3031}
\end{eqnarray}
From these classical constraints we obtain 
\begin{equation}
N_{U_R} + N_{D_R} = N_{J_R} + N_{j_R}
\label{32}
\end{equation}
which through Eq. (\ref{28}) the quantum constraint of Eq. (\ref{27})   
gives a relation between $N$-charges of the exotic quarks      
\begin{equation}
4 N_{J_R} + 5 N_{j_R} = 0
\label{33}
\end{equation}
and from Eq. (\ref{32}) we find 
\begin{equation}
N_{U_R} + N_{D_R} = \frac{1}{5} N_{J_R}.
\label{34}
\end{equation}
If the Standard Model is the input at low energies we know that 
\begin{equation}
N_{U_R} = \frac{2}{3} \quad \mbox{and} \quad N_{D_R} = - \frac{1}{3}
\label{35}
\end{equation}
and then from Eqs. (\ref{33}) and (\ref{34}) we obtain the electric 
charges of the exotic quarks    
\begin{equation}
N_{J_R} = \frac{5}{3} \quad \mbox{and} \quad N_{j_R} = -\frac{4}{3}.
\label{36}
\end{equation}
At this stage it is also possible to establish the last U(1)$_N$ 
charges of the new $G_{331}$ atributions. Let us take the quantum 
constraint of Eq. (\ref{28}) 
\begin{equation}
N_{Q_{1L}} = - 2 N_{Q_{\alpha L}}
\label{37}
\end{equation}
and the cubic quantum constraint of Eq. (\ref{29}) which, in turn,   
may be related to give 
\begin{equation}
N_{Q_{1L}} = \frac{2}{3}
\label{38}
\end{equation}
and 
\begin{equation}
N_{Q_{\alpha L}} = - \frac{1}{3}, \quad \alpha = 2,3
\label{39}
\end{equation}
for the three families of chiral left-handed quarks. 
\par
The $G_{331}$ model is indistinguishable from the Standard Model 
at low energies. In this class of models in order to cancel anomalies 
the number of families, $N_{\rm fam}$, must be divisible by the 
number of colors degrees of freedom, $N_C$. Hence the simplest 
possibility is $N_{\rm fam}/N_C = 1$. Concerning the fermion 
representation content the salient features of 
$G_{331}$ model can be summarized as 1) half the number of 
fermions are put in the SU(3)$_L$ triplet representation and the 
other half in the antitriplet representation; 2) the triangle 
anomalies cancel between families which gives the first step to  
understand the flavor question; 3) the anomaly cancellation  
takes place when the number of families is an integer factor     
of the number of quark colors; 4) a different treatment 
of one quark family than the other two. In particular, a 
singularization of the third family~\cite{quatro} may give us some 
indication as to why the top flavor is so heavy and it may present   
a new approach for the question of fermion mass generation~\cite{doze}; 
5) the existence of new heavy quark flavors at energy scales that 
are higher than those relevant for the Standard Model. For all 
appearences this is a trash but if the cross-section 
$\sigma (p\bar p\rightarrow t\bar t + X)$ obtained by the 
CDF Collaboration~\cite{treze} is in fact higher than the prediction of 
quantum chromodynamics, this may be a signature of new quarks. 
\par
An interesting fact concerns the generalization from SU(3)$_L$ to 
SU(4)$_L$. Using again the lightest leptons as the particles    
which determine the approximate symmetry, if each family is treated 
separately, SU(4) is the highest symmetry group to be considered in 
the electroweak sector~\cite{catorze}. In this sense this is the    
maximal generalization of $G_{331}$ model. There is no room for    
SU(5)$_L \otimes $ U(1) if the nature restrict to the case of 
leptons with $0$, $\pm 1$ electric charges.
\par
From the renormalization group analysis of the gauge coupling 
constants, the breaking scale is estimated to be 1.7 TeV or 
lower~\cite{seis}. We, therefore, expect the masses of dilepton gauge  
bosons and the three flavor exotic quarks to be around or less 
than 1 TeV. The prospects of searching for dilepton gauge bosons 
was considered recently~\cite{quinze} where the cross section for the  
process $e^- p \rightarrow e^+ +\,\,{\rm anything}$ mediated by 
doubly-charged dileptons at HERA with $\sqrt s = $ 314 GeV and an 
integrated luminosity 100 ${\rm pb}^{-1}$ could indicate the 
signature of dileptons with mass up to 340 GeV (650 GeV) if the new 
$j$ quark has a mass lighter than 200 GeV (150 GeV). At LEPII-LHC 
with $\sqrt s = $ 1790 GeV and an anticipated annual luminosity 
6 ${\rm fb}^{-1}{\rm yr}^{-1}$, at least 280          
events per year can be expected  unless both the masses of  
dileptons and of the 
$j$-quark are heavier than 1 TeV. The $j$ or $\bar j$ quarks 
may also be produced in powerful $pp$ colliders such as LHC, 
through the process ${\rm gluon} + {\rm gluon} \rightarrow j + \bar j$.  
The signal for a produced $j$-quark is characterized by 1 jet + 2 leptons,  
since the $j$ quark decays as $j\rightarrow u + l^- + l^-$ through 
$U^{--}$ exchange or as $j\rightarrow d +l^- + \nu_l$ through $V^-$ 
exchange. Indeed, much of the appeal of the $G_{331}$ model is that the 
new physics is guaranteed to be below a few TeV, well within the reach 
of future colliders. Finally, could be that 331 models are not just 
an embedding of the Standard Model but an alternative to describe 
these same interactions and new ones.         

\acknowledgments

I would like to thank the Funda\c c\~ao de Amparo \`a Pesquisa do       
Estado de S\~ao Paulo (FAPESP) for a research fellowship.

\end{document}